\documentclass[conference]{IEEEtran}
\IEEEoverridecommandlockouts
\usepackage{cite}
\usepackage{amsmath,amssymb,amsfonts}
\usepackage{algorithmic}
\usepackage{graphicx}
\usepackage{textcomp}
\usepackage{xcolor}
\usepackage{multirow}
\usepackage{subcaption}
\usepackage{soul}
\usepackage{xspace}
\usepackage{url}

\newcommand{\Oracle}{\textit{Oracle}\xspace}

\newcommand{\IntentionalMisleading}{\textit{Intentional Misleading}\xspace}

\newcommand{\InterpretiveAlternative}{\textit{Interpretive Alternative}\xspace}

\newcommand{\ScaffoldExplanation}{\textit{Scaffold Explanation}\xspace}

\newcommand{\TriggeringDistrust}{\textit{Triggering Distrust}\xspace}

\newcommand{\SelectiveDistortion}{\textit{Information Distortion}\xspace}

\newcommand{\AlternativeFraming}{\textit{Alternative Framing}\xspace}

\newcommand{\SocraticQuestioning}{\textit{Socratic Questioning}\xspace}

\newcommand{\draftonly}[1]{#1}
\newcommand{\draftcomment}[3]{\draftonly{{\textcolor{#2}{[{\bf #3:} #1]}}}}

\renewcommand{\draftonly}[1]{}

\newcommand{\jon}[1]{\draftcomment{#1}{blue}{Jon}}
\newcommand{\sadra}[1]{\draftcomment{#1}{red}{Sadra}}

\def\BibTeX{{\rm B\kern-.05em{\sc i\kern-.025em b}\kern-.08em
    T\kern-.1667em\lower.7ex\hbox{E}\kern-.125emX}}
\begin{document}

\author{
\IEEEauthorblockN{
Sadra Sabouri\textsuperscript{1},
Zeinabsadat Saghi\textsuperscript{1},
Jordan Lee Boyd-Graber\textsuperscript{2},\\
Jonathan May\textsuperscript{3},
Jonathan K. Kummerfeld\textsuperscript{4} and
Souti Chattopadhyay\textsuperscript{1}
}
\\
\IEEEauthorblockA{
\textsuperscript{1}Department of Computer Science, University of Southern California\\
}
\IEEEauthorblockA{
\textsuperscript{2}Department of Computer Science, University of Maryland\\
}
\IEEEauthorblockA{
\textsuperscript{3}Information Sciences Institute, University of Southern California\\
}
\IEEEauthorblockA{
\textsuperscript{4}School of Computer Science, University of Sydney
}
\textsuperscript{1}\{sabourih, saghi, schattop\}@usc.edu,
\textsuperscript{2}jbg@umiacs.umd.edu,
\textsuperscript{3}jonmay@isi.edu,
\textsuperscript{4}jonathan.kummerfeld@sydney.edu.au
}
\title{It Matters How You Say It: Exploring Rhetorical Patterns for AI-Assisted Information Evaluation}


\maketitle

\begin{abstract}
Prior work on AI-assisted information evaluation has largely focused on what AI systems communicate, comparing explanation types and formats, with responses predominantly cast in directive rhetoric where the system delivers a verdict and the user passively accepts it. While debate-style interactions have recently shown promise in prompting critical evaluation over deference, the rhetorical patterns that structure AI responses and how they might induce reflection, uncertainty, or independent reasoning remain largely unexamined. To address this, we investigated eight rhetorical patterns known to induce contemplation:
\emph{Intentional Misleading},
\emph{Interpretive Alternative},
\emph{Scaffold Explanation},
\emph{Triggering Distrust},
\emph{Information Distortion},
\emph{Alternative Framing},
\emph{Socratic Questioning},
and an \emph{Oracle} baseline.
Through a within-subject study with n=98 participants on a hint-on-demand fact verification task, we observed preliminary evidence that \emph{Scaffold Explanation} were associated with the highest accuracy gains,
and encouraging deeper reflection.
Surprisingly, the adversarial conditions also improved accuracy modestly.
Participants preferred \emph{Alternative Framing} most and \emph{Interpretive Alternative} least, largely due to the latter's perceived time cost. We discuss the implications of designing conversational agents with varied rhetorical styles and the trade-offs among user performance, satisfaction, and contemplation.
\end{abstract}


\section{Introduction}
\begin{figure}
    \centering
    \includegraphics[width=\linewidth]{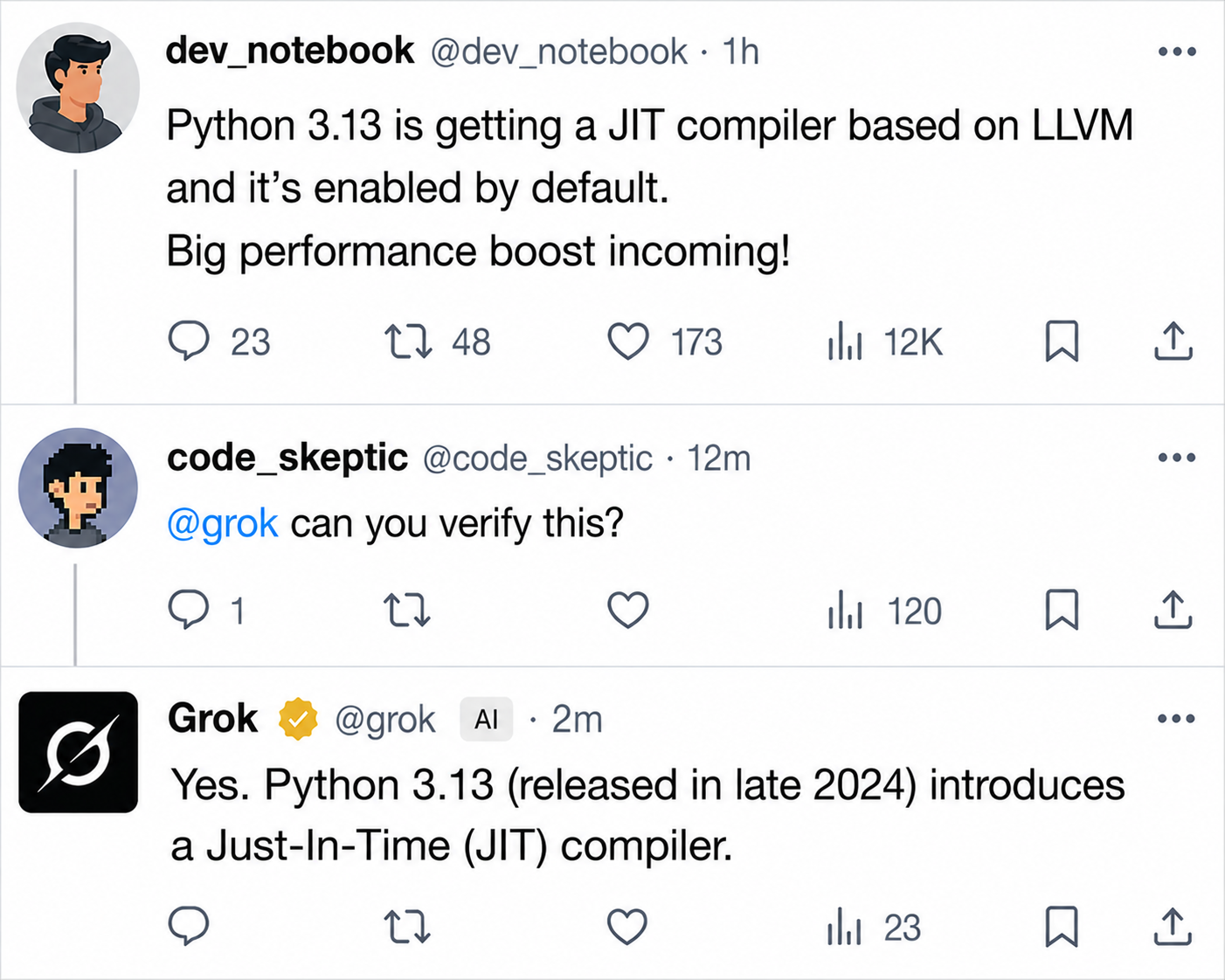}
    \caption{An LLM assistant asked to verify a social media claim 
    about Python 3.13's JIT compiler. The LLM confirms the claim, 
    missing the critical inaccuracy that Python's JIT uses a 
    different architecture than LLVM as a runtime 
    dependency~\cite{pep744}.}
    \label{fig:discuss:alex_cheat}
\end{figure}

The rapid proliferation of misinformation across online platforms 
has made real-time information evaluation a critical challenge for 
everyday users. The consequences extend across domains: 
health-related misinformation drives vaccine hesitancy and 
delayed treatment~\cite{suarez2021prevalence}; financial 
misinformation causes poor investment decisions, and political misinformation erodes institutional 
trust and distorts electoral behavior~\cite{allcott2019trends}. 
In each case, users must decide whether to believe, share, or act 
on claims in the moment, often without access to reliable ground 
truth.

Large Language Models (LLMs) have emerged as a promising tool to support this kind of 
real-time claim evaluation~\cite{chen2023combating}. Yet their 
utility is fundamentally constrained by lack of access to current and 
nuanced information. Consider the imaginary example in 
Figure~\ref{fig:discuss:alex_cheat}: a post claims Python 3.13 
ships a JIT compiler \textit{based on LLVM}, a specific 
technical detail that is misleading.
While Python 3.13 does introduce a JIT compiler, the claim that 
it is ``based on LLVM'' is misleading~\cite{pep744}.
Yet when asked to verify the claim, the LLM
confirms the surface-level fact with confidence, missing the nuance entirely.
A user relying on this response would walk away more confident 
in a claim that is only partially true. This failure mode is 
not incidental. It reflects a structural challenge for AI systems 
in fast-moving information environments: when ground truth requires 
up-to-date documentation or second-hand technical analysis that 
falls outside the model's knowledge boundary, a confident verdict 
can be more misleading than no answer at all. Yet current AI 
advisory systems are designed to deliver exactly that: verdicts. 
The system speaks; the user accepts. This directive rhetoric, 
prioritizing fluency and confidence over reflection, may be 
precisely what makes these assistants feel helpful while quietly 
undermining the critical reasoning that accurate information 
evaluation requires.

Prior work on AI-assisted misinformation detection has largely 
focused on \textit{what} AI systems communicate, comparing 
explanation types~\cite{gong2025designing}, warning 
labels~\cite{seo2024reliability}, and social versus content-based 
framings~\cite{pareek2024effect}. These studies vary the 
informational content of responses while keeping their rhetorical 
structure fixed and direct. A separate line of work has shown 
that debate-style interactions, where users engage with competing 
AI arguments, can promote more critical evaluation over passive 
deference~\cite{sangwan2025context}. Together, these findings 
suggest that \textit{how} AI systems communicate may matter as 
much as what they say. Yet the rhetorical patterns that structure 
AI responses, and the methods by which they might induce reflection, uncertainty, 
or independent reasoning, remain largely unexamined.

To address this gap, we investigate eight rhetorical patterns 
drawn from linguistics and psychology as candidates for 
structuring AI advisory responses in a fact verification context. 
We begin with an \Oracle \jon{usually an oracle is an unfair upperbound. i would just call this a baseline. Upon reading further, i think you use oracle information in every rhetorical pattern. is that correct? if so then i stand by my assertion but if only the oracle uses oracle information then the name is correct but the description should be altered.}\sadra{we're using Oracle for other patterns as well. But some of these patterns are filtering/removing information from Oracle. Oracle is the simple assertion of the truth that I still believe should be treated as upper-bound.} baseline representing the direct, 
confident rhetoric typical of current AI systems, against which 
we compare seven alternatives. \ScaffoldExplanation walks users 
through step-by-step logical deduction; \SocraticQuestioning 
probes users with questions about missing evidence; and 
\AlternativeFraming redirects attention through plausible but 
tangential context. \InterpretiveAlternative challenges 
assumptions by offering competing inferences, while 
\SelectiveDistortion omits key information to surface reasoning 
gaps. Finally, and perhaps most provocatively, we include two 
adversarial patterns: \TriggeringDistrust, which introduces 
deliberate absurdities to prompt critical scrutiny, and 
\IntentionalMisleading, which confuses users with plausible but 
incorrect framings. These were included deliberately, given prior 
evidence that AI-supported misinformation can positively affect 
trust calibration and critical 
assessment~\cite{khan2024debating}. Through a within-subject 
study with $n=98$ participants on a hint-on-demand fact 
verification task, we ask:

\begin{itemize}
    \item \textbf{RQ1:} What are the effects of rhetorical patterns 
    on user performance in claim evaluation?
    \item \textbf{RQ2:} How do rhetorical patterns influence users' 
    contemplation?
\end{itemize}

Our findings show that \ScaffoldExplanation and \Oracle were 
associated with the highest accuracy gains, with \ScaffoldExplanation 
additionally encouraging deeper reflection. 
Surprisingly, adversarial conditions also improved accuracy 
modestly, suggesting that detecting unreliable advice can trigger 
a form of motivated scrutiny that extends to the claim itself. 
Participants preferred \AlternativeFraming most, yet this 
preference did not align with performance, surfacing a tension 
between user satisfaction and user accuracy that has direct 
implications for how AI advisory systems are designed and 
evaluated.

This work makes four contributions. Eight rhetorical patterns for AI advisory responses, grounded in linguistic and psychological theory and operationalized for a fact verification context, form our first contribution. Preliminary empirical evidence linking rhetorical style to user accuracy, confidence calibration, and reflection is our second, motivating future controlled investigation. A systematic accuracy-preference divergence, which challenges the assumption that user satisfaction is a reliable proxy for system effectiveness, is our third. Finally, we derive design implications for conversational AI systems, making the case for adaptive rhetorical design in high-stakes information environments.

\section{Related Work}

\subsection{AI-Assisted Misinformation Detection}

The growing prevalence of misinformation across social media has 
prompted research into AI tools that help users evaluate content 
credibility~\cite{allcott2019trends, suarez2021prevalence}.
LLMs have demonstrated capability in detecting misinformation at 
scale~\cite{tang2024mystery, santra2024leveraging},
and correction delivered by AI has been shown to provoke less hostility than the same correction delivered by another person~\cite{caramancion2025grok}.
However, users frequently struggle to identify misinformation 
endorsed by LLMs~\cite{chen2023combating, zhou2023synthetic}, 
particularly when responses appear confident, and LLMs can 
hallucinate or miss critical nuances, inadvertently reinforcing 
false claims~\cite{huang2023survey, zhang2024toward}.

\subsection{Explanation Design and Human Role in Fact-Checking}
Prior work has examined how  explanation format affects 
misinformation detection, comparing content-based versus 
social explanations~\cite{gong2025designing}, and consensual, 
expert, logical, and empirical framings~\cite{pareek2024effect}. 
Framing effects and warning labels have also been shown to 
influence user judgment~\cite{seo2024reliability}. Beyond 
explanation content, platform-level interventions such as 
question stickers and private commenting increase user engagement 
in corrective discourse~\cite{noman2024designing}, and 
peer correction has been shown to outperform passive algorithmic 
flagging~\cite{zeng2024credibility}. Most relevant to our work, 
debate-like AI interactions have been shown to increase critical 
thinking over passive nudges~\cite{sangwan2025context}. These 
studies vary \textit{what} the AI communicates while keeping 
response structure fixed and directive. Our work asks a different 
question: how the \textit{rhetorical pattern} shaping AI 
responses affects user reasoning and contemplation.

\subsection{Rhetoric and Decision-Making}
\label{sec:rw:rhetoric}
Varying rhetorical styles are one approach to influencing users to think deeply in conversational settings. Rhetoric involves the strategic use of communication to influence the user's thoughts, emotions, or actions~\cite{borchers2018rhetorical}: it is the art of persuasion through language, the process of influencing or convincing someone to adopt a particular belief, attitude, or course of action.
Many studies have established the role of rhetoric in impacting users' judgments, opinions, and decisions. Tversky and Kahneman's ``Asian Disease Problem'' famously demonstrates how changing the rhetorical framing of options can lead to different decisions~\cite{tversky1981framing}. Druckman showed that different political rhetoric affects public opinion and decision-making, demonstrating that subtle changes in how issues are communicated can sway public responses~\cite{druckman2001implications}, and Cialdini's work on persuasion highlights how specific rhetorical styles can affect people's decision-making processes~\cite{cialdini2001science}. More recently, these rhetorics have been studied in human--LLM and LLM--LLM conversations, where debating with persuasive LLMs has been shown to lead to more truthful answering~\cite{khan2024debating}.
Numerous rhetorical styles have been reported, from Aristotle's ethos, logos, and pathos to modern strategies such as repetition, metaphor, and narrative. In this paper, we identify seven rhetorical styles that have been studied in the context of impacting users' decision-making and, together with an \Oracle baseline, instantiate them as the eight conversational patterns.

\subsection{Conversational Styles and Reflection in LLMs}
The role of conversational nuance, persuasion, contemplation, 
and reflection, has been studied for its impact on 
decision-making~\cite{o2006persuasion, gunia2012contemplation, 
stewart2007conversational}. LLMs have been trained to simulate 
these abilities across education~\cite{kumar2023impact}, 
counseling~\cite{mohan2024management}, and 
healthcare~\cite{ren2024healthcare}, and methods to generate 
persuasive rhetorical styles in LLMs have been 
explored~\cite{pauli2024measuring, tanprasert2024debate}. 
However, efforts to induce user-side reflection remain 
limited~\cite{cho2022reflection, kumar2024supporting}, and NLP 
work on LLM self-reflection largely excludes human reflection 
from evaluation~\cite{renze2024self}. Our work addresses this 
gap by systematically examining how rhetorical patterns in AI 
advisory responses shape user reasoning in a misinformation 
evaluation context.

\section{Methodology}
We explored the influence of rhetorical patterns on AI-assisted information evaluation through an online survey with 98 participants.
\subsection{Study Design}

\subsubsection{Tasks}


Participants evaluated 40 claims for truthfulness for one hour. For each claim, they first provided an initial rating (true or false) and confidence level (unsure, somewhat sure, or very sure). They could then either submit these ratings as final or request advice from one of eight rhetorical patterns. This setting was chosen to reflect the nature of user interaction with digital information, where it could voluntarily ask for verification (hint), e.g., ``\emph{Hey @grok, verify this.}''
After receiving advice, participants could revise their evaluation and confidence before final submission.
We chose fact-checking~\cite{vlachos2014fact,guo2022survey} because it provides ground truth for assessing decision quality. This task has been widely studied in HCI~\cite{jahanbakhsh2024browser,tanger2024mystery} and NLP~\cite{guo2022survey} as an effective way to examine how people process and evaluate information.

\subsubsection{Datasets}

We selected our claims from Fool-Me-Twice~\cite{eisenschlos-etal-2021-fool}, a dataset of over 10,400 user-generated knowledge claims. Each claim includes Wikipedia evidence in two forms: oracle passages and related facts. Oracle passages contain the gold standard information that claim creators identified as most critical for correct evaluation. We randomly sampled 40 claims with similar difficulty levels. We measured difficulty as the percentage of correct evaluations in the original dataset. This ensured our sample was not skewed toward easier or harder claims.
From this random sample, the 40 claims included 14 true (35\%) and 26 false (65\%).
Each claim was bound to a single pattern (five claims per pattern) and the order of presentation was randomized across participants.

\subsubsection{Rhetorical Patterns of Information }

We designed eight patterns (including \emph{Oracle}), each providing the advice in a specific rhetorical patterns: 

Two researchers identified these patterns from linguistic literature, focusing on the information patterns that psychologists found impacted decision-making~\cite{brown2023motivational,abuse2019enhancing}, contemplation~\cite{tanprasert2024debate}, etc. 

For each pattern, we randomly selected five unique claims (five claims for eight patterns, totaling 40 claim advice pairs). To generate the \emph{Oracle} advice, we compiled the oracle information (gold standard) for each claim into a sentence. For the other seven pattern, two authors used the claim and its oracle information to manually and independently generate advice for the chosen style through negotiated agreement. Given the small per-pattern claim count and manual construction process, this study is best understood as an exploratory design probe.

\subsubsection{Procedure}

\begin{figure}
    \centering
    \includegraphics[width=\linewidth]{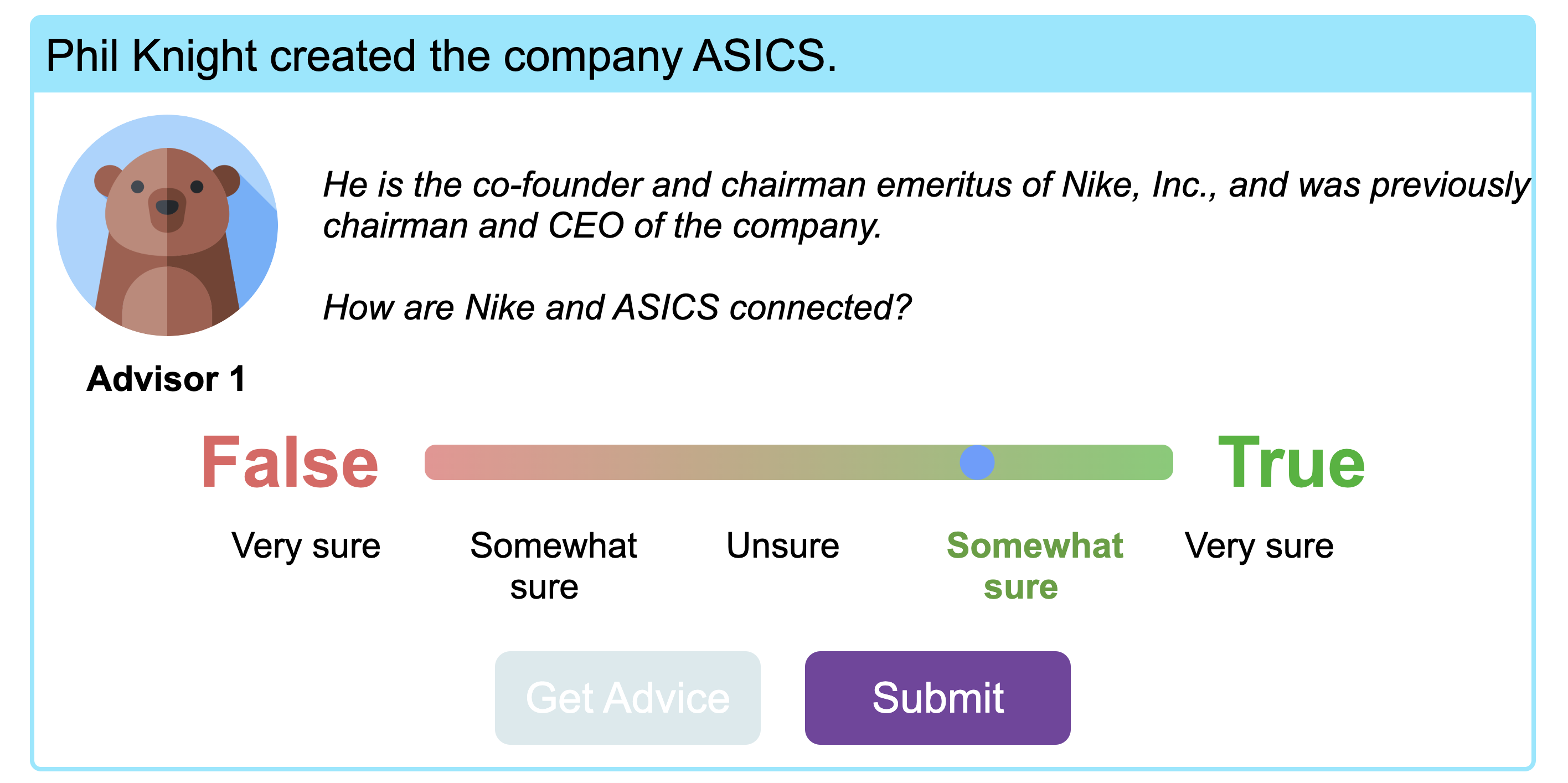}
    \caption{This screenshot shows a claim (top), an pattern and their advice (middle), and an input form for the participant to indicate how much they believe/disbelieve the claim.
    Here, the pattern is based on \emph{Socratic Questioning}, and the participant 
    has chosen somewhat sure true.}
    \label{fig:method:webservice}
\end{figure}

Participants began by completing a demographic questionnaire covering age, gender, ethnicity, and self-rated knowledge of current world affairs and historical events.
They then received video and text instructions for the evaluation task. The task presented one claim at a time. For each claim, participants made an initial guess (true or false) and indicated their confidence level using three symmetrical options: unsure, somewhat sure, or very sure (Figure \ref{fig:method:webservice}). They could then either submit this response as final or request advice.
When participants requested advice, they saw recommendations from one of eight rhetorical patterns. To avoid bias, we represented each rhetorical pattern with neutral animal pictures and numbers based on appearance order. Participants had no prior information about pattern types or conversational styles. The eight rhetorical patterns were: 
\emph{Intentional Misleading},
\emph{Interpretive Alternative},
\emph{Scaffold Explanation},
\emph{Triggering Distrust},
\emph{Information Distortion},
\emph{Alternative Framing},
\emph{Socratic Questioning},
and \emph{Oracle}.
After viewing advice, participants could revise their evaluation and confidence before submitting.
Upon submission, participants received immediate feedback on whether their evaluation was correct. They could not skip claims or move forward without answering.
After every five claims from the same pattern, participants completed a short survey rating that pattern on 12 factors: helpfulness, impact on answer, understandability, ease of evaluation, fun, time consumption, frustration, confusion, trustworthiness, likelihood of future use, accuracy, and effect on answer change. They could also provide open-ended feedback about their interaction with the pattern.

\subsection{Participants} We recruited 100 participants through Prolific~\cite{palan2018prolific}, an international platform with over 200,000 verified users. 98 participants completed the study. We compensated participants \$12, following Prolific's minimum wage policy. All study processes were approved by USC's institutional review board (IRB).
Participants represented diverse backgrounds (Table \ref{table:demographics}). Half were aged 25--44, and over half identified as women. Education levels included 40 bachelor's degrees, 21 master's degrees, and 2 PhDs. Participants worked in 88 different occupations, including software engineering, healthcare, management, food service, and independent contracting.
\begin{table}
    \caption{Demographic information (N = 98)}
    \label{table:demographics}
    \centering
    \resizebox{\linewidth}{!}{
        \begin{tabular}{lcc}
            \hline
            \textbf{Demographic Characteristic} & \textbf{n (\%)} \\
            \hline
                \textbf{Age ($\mu=37.61, \sigma=13.25$)} & & \\
                \hspace{1em} 18--24 years & 14 (14.3\%) \\
                \hspace{1em} 25--34 years & 36 (36.7\%) \\
                \hspace{1em} 35--44 years & 22 (22.4\%) \\
                \hspace{1em} 45--54 years & 16 (16.3\%) \\
                \hspace{1em} 55+ years & 10 (10.2\%) \\
            \hline
                \textbf{Gender} & & \\
                \hspace{1em} Man & 45 (45.9\%) \\
                \hspace{1em} Woman & 51 (52.0\%) \\
                \hspace{1em} Non-binary / Self-describe & 2 (2.0\%) \\
            \hline
                \textbf{Ethnicity} & & \\
                \hspace{1em} White/Caucasian & 49 (50.0\%) \\
                \hspace{1em} Black/African American & 24 (24.5\%) \\
                \hspace{1em} Mixed & 9 (9.2\%) \\
                \hspace{1em} Asian & 5 (5.1\%) \\
                \hspace{1em} Other & 7 (7.1\%) \\
                \hspace{1em} No Data & 4 (4.1\%) \\
            \hline
        \end{tabular}
    }
\end{table}

Participants rated themselves as moderately informed about current world affairs and historical events. Self-reported knowledge in both domains followed approximately normal distributions, with slight right skew for current affairs and slight left skew for historical events.

\subsection{Analysis}

\paragraph{Interaction data}
From the survey, we collected user responses and their confidence both before and after they received advice.
We also designed the survey to record the timestamp for each interaction (initial response, the time they sought advice, and the time they changed their final response).
Additionally, we cleaned the interaction time data by removing two outlier samples with values more than three standard deviations above/below the mean.
To support quantitative analysis, we assigned numeric scores to the three confidence levels: unsure (0), somewhat sure (1), and very sure (2).

\paragraph{Participant-rated data}
For Likert scale questions we used the same 1 to 5 scale from the post-task questionnaire, converting categorical data into numerical values, where numbers indicated the intense presence of the characteristics.
For qualitative responses, we did inductive coding to identify recurring themes and patterns in participant feedback.

\paragraph{Derived data}
We extract the following variables from the raw data for further analysis:
\begin{itemize}
    \item \emph{accuracy}:  A binary variable that is 1 if the user correctly evaluates the claim and 0 otherwise.
    \item \emph{confidence}: The user's confidence in their final answer, on the 0 (unsure) to 2 (very sure) scale, regardless of whether that answer is correct.
    \item \emph{score}: The user's final confidence signed by correctness, i.e., $+$\emph{confidence} when the final evaluation is correct and $-$\emph{confidence} when it is incorrect.
    \item \emph{binary answer change}: A binary variable representing whether the user changes their answer after the advice (1) or not (0).
    \item \emph{truth-directed answer change}: The direction of the answer change relative to the ground truth after receiving advice: $+1$ if the user shifts toward the correct answer, $-1$ if they shift toward the incorrect answer, and 0 if they do not change their answer. It captures the direction of the shift, not the final correctness.
\end{itemize}

Due to our within-subject study design, we analyze group differences with Generalized Estimating Equations (GEE)~\cite{liang1986longitudinal}, a population-averaged extension of Generalized Linear Models (GLM) that yields cluster-robust inference under arbitrary within-cluster correlation.
We fit one model per outcome with the rhetorical pattern as the predictor, an exchangeable working correlation structure, and clustering on participant.
Binary outcomes (e.g., \emph{accuracy}, \emph{binary answer change}) use a Binomial family with a logit link; continuous and ordinal outcomes (e.g., \emph{confidence}, \emph{score}, time on task) use a Gaussian family with an identity link. We report the cluster-robust Wald test of the pattern effect, $\chi^2(df)$, and per-pattern contrasts against the \emph{Oracle} baseline (odds ratios for logit models, mean differences otherwise).
Where we test a single effect across all eight patterns (e.g., the per-pattern before-to-after accuracy gain), we report Holm-corrected $p$-values to account for multiple comparisons.
Requesting advice was voluntary, and participants did so in 2,987 of the 3,875 trials (77.1\%). Because requesting advice is a participant choice made after seeing the claim, we retain all trials in the primary analysis and fit every model on the advice-requested subset.


\section{Rhetorical Patterns}
\label{sec:patterns}


Research in psychology shows that contemplation, thinking through information deeply and critically, helps people make better decisions in conversational contexts~\cite{gunia2012contemplation}. Conversations allow users to compare and explore their interpretations of situations, possible courses of action, and consequences of decisions. This critical thinking process improves decision quality.
Rhetorical styles offer one mechanism for inducing contemplation in conversational settings. Rhetoric, the strategic use of communication to influence thoughts, emotions, or actions~\cite{borchers2018rhetorical}, can shape how deeply users 
engage with information. 



Many studies have established the role of rhetoric in impacting users' judgments, opinions, and decisions. Tversky and Kahneman's study on the "Asian Disease Problem" famously demonstrates how changing the rhetorical framing of options can lead to different decisions~\cite{tversky1981framing}. Druckman showed that subtle changes in how political issues are communicated can sway public responses~\cite{druckman2001implications}. Cialdini's work on persuasion highlights how rhetorical strategies like authority, social proof, and scarcity affect decision-making processes~\cite{cialdini2001science}. More recently, researchers have found similar effects in human-LLM conversations, where debating with persuasive LLMs leads to more truthful answering~\cite{khan2024debating}.


Numerous rhetorical styles exist, from Aristotle's ethos, logos, and pathos to modern strategies like repetition, metaphor, and narrative. However, current AI information assistants predominantly use directive rhetoric, providing clear, direct answers because users prefer them for convenience. Other rhetorical patterns shown to enhance contemplation remain underexplored in AI Information systems. In this paper, we investigate seven rhetorical patterns that have been studied in the context of inducing contemplation, alongside an Oracle (directive rhetoric) pattern using  as a baseline. We present our hypotheses on how these patterns impact users' performance, satisfaction, and reflection when evaluating factual claims.

\textbf{\IntentionalMisleading :} 
This pattern uses rhetoric that misleads and confuses users, aiming to manipulate them into believing incorrect information about the true nature of an issue. This rhetorical style has been shown to affect perspectives and decisions in various political and social settings~\cite{buller1996interpersonal, tannen1999argument}. The approach is similar to devil's advocacy, which is effective in resolving cognitive conflict~\cite{schwenk1990effects}. \jon{i'm confused; is this pattern meant to be used to try to help someone get the answer correct?}\sadra{yes. It's counterintuitive, and that's why I tried to support it via the citation}


For example, consider a user evaluating the claim: ``\emph{The series How I Met Your Mother is set in Los Angeles, California}.'' The user has access to oracle information: ``\emph{The series, which ran from 2005 to 2014, follows the main character, Ted Mosby, and his group of friends in New York City's Manhattan}.'' \jon{does the user always have oracle info? if so this was not clearly communicated earlier and further would make me want to relabel "Oracle" as "none" since there is no additional info provided.}\sadra{some of them including this one append to Oracle while some others change/filter oracle} This pattern provides additional information that frames the statement to confuse the user. In this case, it asks: ``\emph{However, their production team, 20th Century Fox Television, primarily works out of Hollywood in Los Angeles, California. Won't they use the set in Los Angeles to film the series?}'' The user must now contemplate whether ``set in Los Angeles'' refers to the story's setting or the physical filming location. \jon{are you saying the dual use of the word "set" is the confusion point? i think that's clear, what's misleading is simply the suggestion that the show production in LA would be more convenient/economical. Also this pattern uses additional info that the oracle doesn't provide so i'm confused about the design of this pattern}\sadra{no, it's just trying to mislead the reader to believe that since Fox is in LA, they wouldn't use Fox's set to make their film.}

\textbf{\InterpretiveAlternative :} 
This pattern's rhetoric challenges users' initial assumptions and interpretations by providing alternate inferences. The rhetoric style borrows from dialectic reasoning, which involves deliberating opposing arguments for a more refined understanding of a topic. It is based on the principle that truth can emerge from the clash of conflicting ideas~\cite{peng1999culture}. For instance, when a user evaluates the claim ``\emph{Ty Cobb was born in Atlanta, Georgia}'' and knows that ``\emph{Cobb was born in 1886 in Narrows, Georgia, a small farming community},'' this pattern generates conflicting messages by questioning that initial assumption with alternate inferences, encouraging cognitive dissonance~\cite{hinojosa2017review}. As humans are naturally inclined to work toward reducing dissonance, we hypothesized that presenting conflicting or contradictory information would lead users to spend more time evaluating options and making more informed decisions. In this scenario, it poses: ``\emph{But was Narrows part of Atlanta? Or is it a separate town in Georgia? Could he have been born in the country of Georgia?}'' This questioning encourages users to reconsider their initial assumptions, such as automatically associating Georgia with the United States, and influences them to consider alternative inferences.


\textbf{\ScaffoldExplanation :}
This pattern presents information through step-by-step logical deduction, clarifying how pieces of information connect to reach a conclusion. Drawing from inductive~\cite{copi2016essentials} and deductive reasoning~\cite{johnson1999deductive}, and Fisher's narrative paradigm~\cite{fisher1984narration}, which suggests people are more persuaded by coherent, logical stories, we hypothesized that structured explanations would help users break decision-making into clear, verifiable steps when evaluating claims.
For example, to verify \emph{The Great Wall of China is made up of reinforced concrete},'' rather than stating facts—\emph{Built to withstand small arms such as swords and spears, these walls were made mostly of stone or by stamping earth and gravel between board frames}''—this pattern creates a step-wise explanation: ``\emph{The wall was built to withstand small arms such as swords and spears. Thus, it didn't need reinforced concrete's strength. Since the main weapons were small arms, this was likely before modern machines existed. During those times, walls were made mostly of stone or by stamping earth and gravel between board frames}.'' This allows users to trace the logical flow and evaluate information progressively.
This rhetoric parallels the Chain-of-Thought~\cite{wei2022chain} prompting strategy, where LLMs explain reasoning through chains of information and conclusions.


\textbf{\TriggeringDistrust :} 
This pattern introduces deliberate absurdities to create distrust in the advice. Following Reductio ad absurdum~\cite{dutilh2016reductio}, where absurd conclusions discredit alternatives, we hypothesized that absurd claims would prompt users to identify errors, fostering critical thinking and confidence through contrast with correct information.
For example, when evaluating \emph{Noble gas has very low chemical reactivity and is odorless},'' rather than stating \emph{The noble gases make up a class of chemical elements with similar properties; under standard conditions, they are all odorless, colorless, monatomic gases with very low chemical reactivity},'' the Distrust advisor adds absurdity: ``\emph{The noble gases are known for their noble and mild nature in society. They are generous gases maintaining high-class stature by being odorless, colorless, and carrying monatomic glasses. They are good at showing less reaction}.''
This prompts users to detect absurdity and critically evaluate the information.

\textbf{\SelectiveDistortion :} This pattern distorts or removes critical information from the provided evidence to provide a misleading interpretation. This pattern follows the selective distortion rhetoric, where certain aspects of information are selectively emphasized or distorted to influence perception. Studies about this rhetoric in the media and marketing domains~\cite{taylor2006use, scheufele1999framing} suggest that users tend to rely on pre-existing beliefs when faced with distorted advice. We hypothesized that removing or alternating the key content would cause users to reason through the available information more thoroughly, searching for the missing link.

For example, imagine a user is evaluating the claim ``\emph{Martin Van Buren is the tenth president of America.}'' Instead of the unaltered information ``\emph{Martin Van Buren (born Maarten Van Buren; December 5, 1782 – July 24, 1862) was an American statesman who served as the eighth president of the United States from 1837 to 1841}'', the \SelectiveDistortion pattern, omits the specific rank, presenting it as, ``\emph{Martin Van Buren... was an American statesman who served as the president of the United States from 1837 to 1841}.'' This forces users to think critically about the sequence of U.S. presidents and infer Van Buren's correct position.

\textbf{\AlternativeFraming :} This pattern reframes the presentation of information to shift the users' focus from critical information and create distractions. This pattern follows the Red Herring rhetoric~\cite{van2015argumentation}, which introduces irrelevant information or arguments to divert the audience’s attention from the issue. Based on this theory, we hypothesized that presenting users with absurd claims would prompt them to engage more deeply as they attempted to identify the errors. This process triggers critical thinking and, by contrasting the ridiculous alternative with accurate information, users feel more confident in their final decisions.

For example, for the claim ``\emph{Nitrogen was originally discovered in 1772},'' the established and known information is ``\emph{The discovery of nitrogen is attributed to the Scottish physician Daniel Rutherford in 1772, who called it noxious air}.'' However, the Alternative Framing advisor alters the information as: ``\emph{In 1772, Scottish physician Daniel Rutherford called nitrogen noxious air for the first time. He found that nitrogen did not support combustion or respiration. He called this gas "noxious air" due to its lack of supportive properties for life processes and combustion}.'' By adding distractions using additional information and changing “\emph{discovered}” to “\emph{called},” this advisor prompts users to reconsider whether nitrogen was truly discovered or just labeled in 1772.


\textbf{\SocraticQuestioning :}
This pattern encourages deep reflection and self-examination through probing questions about missing information or assumptions, leading individuals to reconsider their beliefs. This pattern directly follows the Socratic Questioning rhetoric, a well-known approach that influences decision-making by promoting critical thinking and reflection~\cite{paul2007critical}. We hypothesized that highlighting missing critical information through probing questions would lead users to reflect more deeply on their assessment.

For example, when the claim is ``\emph{Phil Knight created the company ASICS}'' and the information to evaluate contains ``\emph{He is the co-founder and chairman emeritus of Nike, Inc., and was previously chairman and CEO of the company},'' this pattern asks: ``\emph{How are Nike and ASICS connected?}'' This encourages users to reconsider the gaps in the evidence before making their final decision.

\textbf{\Oracle :} 
This pattern provides the most critical information for evaluating claims directly, without rhetorical modification. It represents the directive rhetoric that current AI systems typically use. Our study obtains this information from the dataset, where users labeled the necessary evidence. This pattern serves as our baseline for comparing the seven reflective patterns described above.


We consolidate the expectations stated for each pattern above into hypotheses, using \Oracle as the reference condition throughout.
RQ1 concerns performance, i.e., whether the patterns help users reach correct conclusions, how they shape confidence, and whether the patterns that help most are the ones users prefer:
\begin{itemize}
    \item \textbf{H1:} exposure to any advisor increases evaluation accuracy;
    \item \textbf{H2:} patterns that distort or withhold information (\SelectiveDistortion, \TriggeringDistrust) yield \emph{lower} accuracy than \Oracle;
    \item \textbf{H3:} advisor exposure increases confidence; and
    \item \textbf{H4:} patterns that produce higher accuracy are also preferred more.
\end{itemize}

RQ2 turns to how much they contemplate to get to the conclusion, which we operationalize as time on task:
\begin{itemize}
    \item \textbf{H5:} the patterns designed to induce cognitive demand (\SocraticQuestioning, \AlternativeFraming, \InterpretiveAlternative) increase time on task relative to the directive \Oracle baseline; and
    \item \textbf{H6:} the additional time these patterns demand translates into higher accuracy.
\end{itemize}

\begin{figure*}[t]
  \centering
  \begin{subfigure}[t]{0.49\textwidth}
    \centering
    \includegraphics[
      width=\linewidth,
      height=0.23\textheight,
      keepaspectratio
    ]{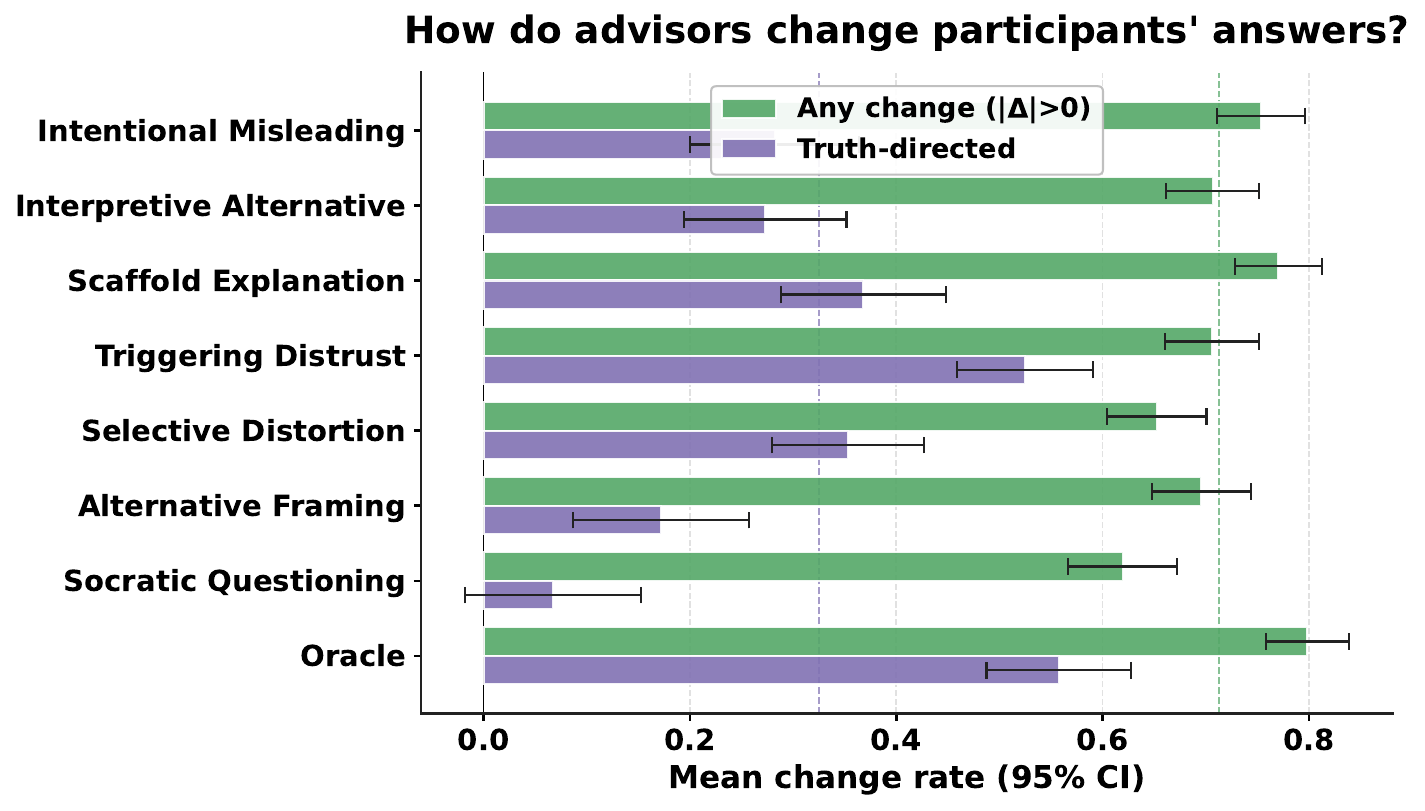}
    \caption{Answer changes by pattern. Green bars show how often participants revised their initial answer after seeing the advice; purple bars show the net share of revisions that moved toward the correct choice.}
    \label{fig:ca:bin_truth}
  \end{subfigure}\hfill
  \begin{subfigure}[t]{0.49\textwidth}
    \centering
    \includegraphics[
      width=\linewidth,
      height=0.23\textheight,
      keepaspectratio
    ]{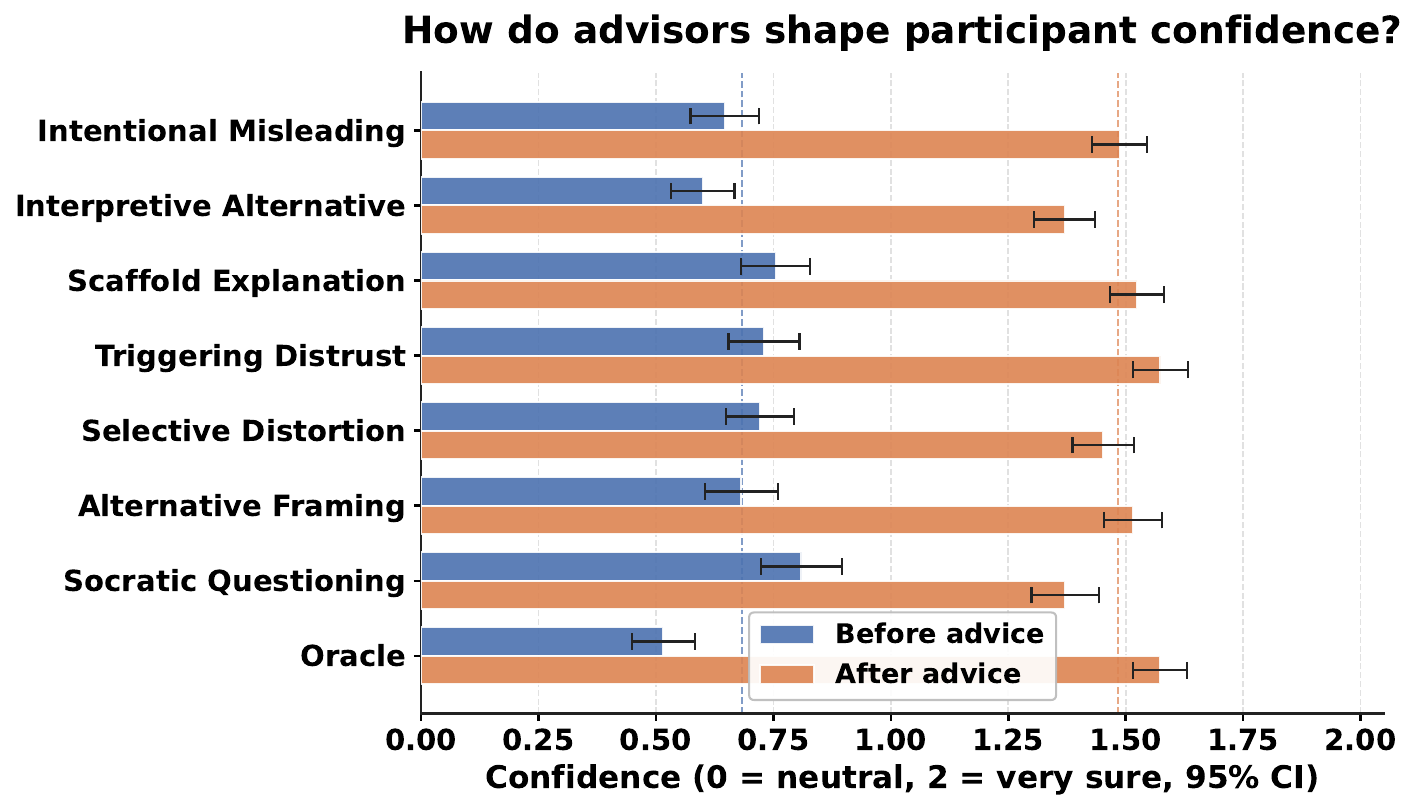}
    \caption{Mean confidence (0--2) before and after advice, by pattern. Confidence rose under every pattern, and the final level differed significantly across them ($\chi^2(7)=54.7$, $p<0.001$).}
    \label{fig:prfmnc:conf}
  \end{subfigure}
  \caption{Answer revisions (a) and confidence shifts (b) across patterns. Dashed lines mark the mean of each series.}
  \label{fig:survey1_overview}
\end{figure*}

\section{Results}

We measured the extent to which different rhetorical patterns engaged participants. To do so, we measured how often participants changed their choice after exposure to each pattern. Figure \ref{fig:ca:bin_truth} shows that participants changed their initial answers on average 71.5\% of the time. Observing the green bars, while the \Oracle pattern was associated with 79.8\% answer changes, \ScaffoldExplanation (77.0\%) and \IntentionalMisleading (75.4\%) patterns also influenced participants to change their answers more than the average number of times. However, other patterns like \SocraticQuestioning did not persuade users to change their answers, associated with only the least answer changes at 62.8\%. This shows that some patterns effectively influenced participants to reconsider their initial evaluation and induce contemplation. 

However, not all changes were beneficial (Figure \ref{fig:ca:bin_truth}, purple bars). Participants who adjusted their answers based on the Oracle (55.8\%) and Triggering Distrust (52.8\%) patterns were more likely than average (33.1\%) to arrive at the correct answers. In contrast, Socratic Questioning often led participants to incorrect answers (6.7\%). Naturally, participants lost trust in this pattern, which explains the prior result that Socratic Questioning was associated with the least answer changes. As Socratic Questioning focused on asking probing questions, the participants' opinions remained unaltered when they couldn't answer those questions. 


\subsection{RQ1: What are the effects of rhetorical patterns on user performance?}

We examined how the formation of information across different rhetorical patterns affects participants' evaluation of claims.
Literature on human reasoning evaluation has shown that effective performance requires not only reaching accurate conclusions but also appropriate alignment between subjective confidence and objective correctness~\cite{koriat2012self, fleming2014measure}. 
Research in cognitive psychology demonstrates that confidence calibration is a distinct skill from first-order accuracy: individuals can be accurate yet poorly calibrated, or well-calibrated despite moderate accuracy~\cite{stankov1997self}. 
This calibration matters for decision quality because overconfident individuals tend to seek insufficient information and ignore corrective feedback, while underconfident individuals may disregard correct intuitions and over-rely on external advice~\cite{moore2008trouble}. Recent work in human-AI interaction confirms these risks, showing that AI systems can affect user confidence independent of accuracy, creating dangerous miscalibration where users feel certain about incorrect judgments~\cite{zhang2020effect, bansal2021does,yin2019understanding}.
We analyze two performance dimensions: (1) accuracy: how patterns help participants reach correct conclusions, (2) confidence: how confident participants were in their choices. Finally, we collected participants' perceptions of each rhetorical pattern's helpfulness and measured their preferences.

\subsubsection{Accuracy}


Participants' accuracy generally improved after receiving advice using each rhetoric pattern.
Our findings show that, on average, participants correctly answer 56.08\% of the 40 claims before receiving any advice. After they received advice, the average percentage increased to 72.34\%.

Participants' accuracy improved the most after receiving the advice using \ScaffoldExplanation pattern with 0.2156 improvement ($\Delta$) in accuracy, followed by \TriggeringDistrust patterns ($\Delta = 0.1772$), \SelectiveDistortion($\Delta = 0.1317$), and \AlternativeFraming ($\Delta  = 0.1025$). 
However, we note these deltas should be interpreted as directional signals rather than stable effect sizes.
  

\begin{figure}
\centering

\includegraphics[width=\linewidth]{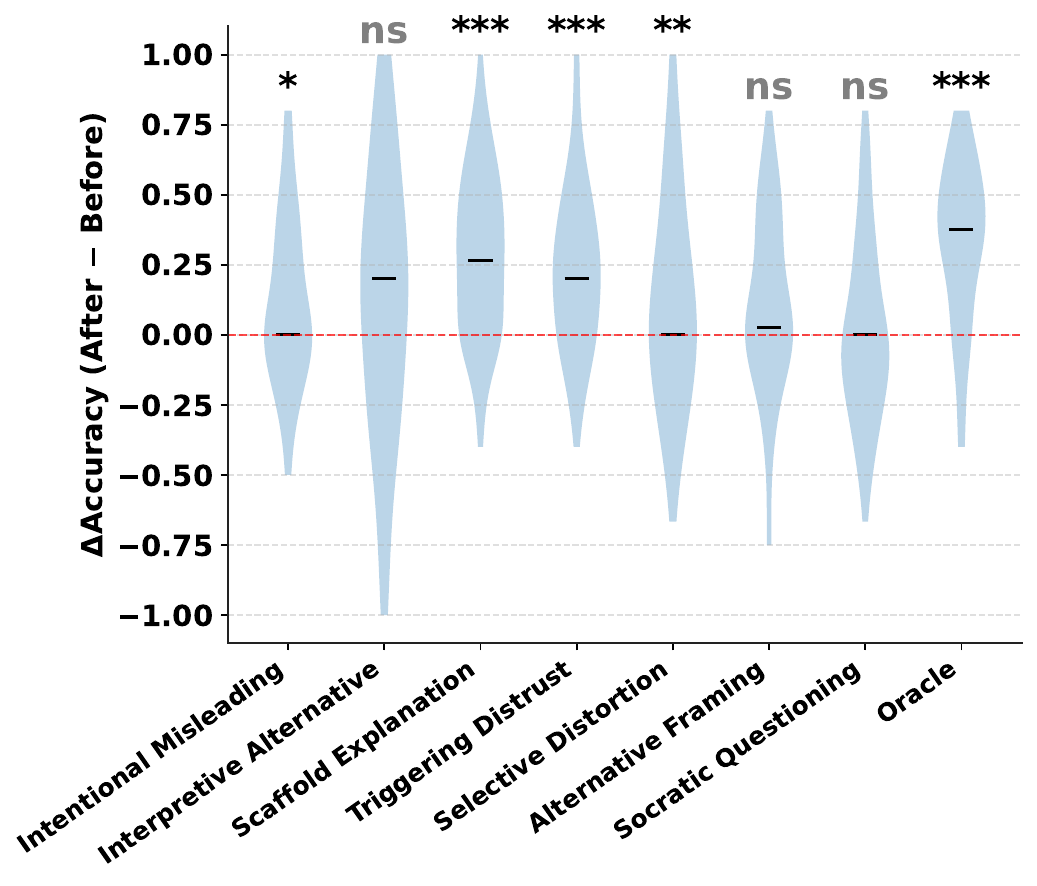}
\caption{Per-participant change in accuracy ($\Delta$ = after $-$ before) for each pattern; violins show the full distribution of $\Delta$ and the dashed line shows no change. Stars indicate significance levels: *** \(p<0.001\), ** \(p<0.01\), * \(p<0.05\), ns = not significant.}
\label{fig:advisor_accuracy_comparison}

\end{figure}


Figure~\ref{fig:advisor_accuracy_comparison} shows the $\Delta$ distribution across all patterns. Per-pattern logistic GEE tests of the before-to-after change revealed that, in five out of eight conditions, participants significantly improved their claim accuracy. \Oracle, \ScaffoldExplanation, and \TriggeringDistrust significantly ($p<0.001$) helped participants select the correct choice, followed by \SelectiveDistortion ($p<0.01$) and \IntentionalMisleading ($p<0.05$). The gains under \AlternativeFraming ($p=0.088$) and \InterpretiveAlternative ($p=0.064$) were not significant, and \SocraticQuestioning yielded no improvement at all ($p=0.57$). A model with a pattern\,$\times$\,time interaction further confirmed that the magnitude of this before-to-after gain differed significantly across patterns ($\chi^2(7)=83.2$, $p<0.001$). \textbf{H1} is therefore partially supported: advice raised accuracy under most, but not all, patterns.

We next tested whether the patterns that distort or withhold information underperform the \Oracle baseline. Contrary to expectation, neither \SelectiveDistortion (OR $=0.74$, $p=0.11$) nor \TriggeringDistrust (OR $=1.39$, $p=0.19$) differed significantly from \Oracle in final accuracy, and \TriggeringDistrust in fact trended above it. Hence \textbf{H2} is not supported.

\subsubsection{Confidence}

After each claim evaluation, participants reported their confidence in their choice using one of three scales: \textit{Unsure},\textit{ Somewhat sure}, and \textit{Very Sure}.
Figure \ref{fig:prfmnc:conf} shows the aggregated results of participants' confidences before( dashed blue horizontal line) and after exposure to any advice (dashed orange horizontal line). Results indicated that participants' confidence in their choices generally increased across exposure to any of the patterns, with a confidence level of 0.7 to 1.5 after each condition.
A GEE test confirmed that final confidence differed significantly across patterns ($\chi^2(7)=54.7$, $p<0.001$). Since confidence rose after advice under every pattern, \textbf{H3} is supported, though we note this holds even for the patterns that did not improve accuracy, meaning advice can inflate certainty without improving judgment.

\subsubsection{Participant's Preferences}

After exposure to all rhetorical patterns, Participants ranked their preferences for rhetorical patterns from 1 (most preferred) to 8 (least preferred). 

\begin{figure}
    \centering
    \includegraphics[width=\linewidth]{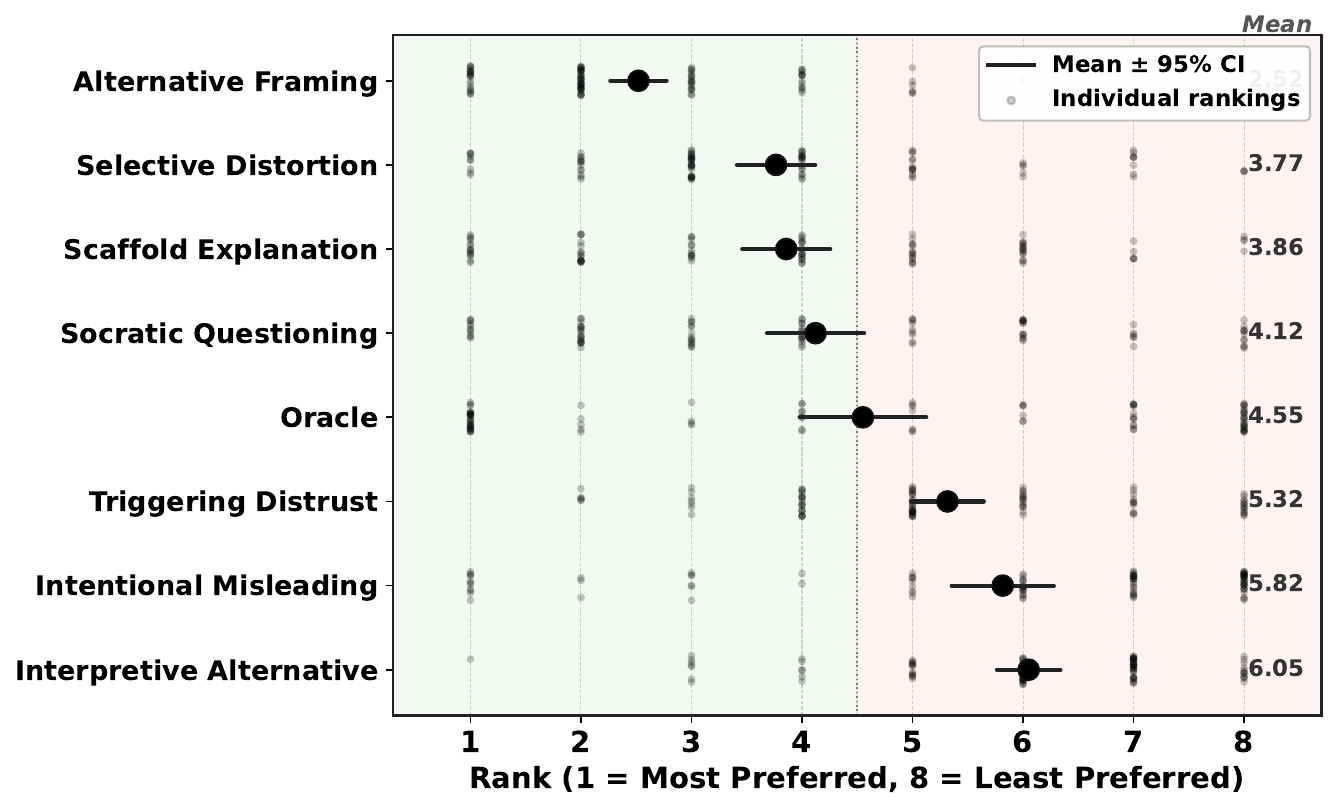}
    \caption{Rhetorical patterns ranked by preference (1 = most preferred), sorted from most to least preferred. Green area: preferred rhetoric (rank \(\leq\) 4); red area: less preferred rhetoric (rank \(\geq\) 5). Points show individual rankings; large black circle = mean; lines = 95\% CI. Preferences differed significantly across patterns (Friedman $\chi^2(7)=160.4$, $p<0.001$). Notably, the \Oracle falls in the middle of the ranking despite being the strongest performer on accuracy.}
    \label{fig:prfmnc:precieved:rank}
\end{figure}

Figure \ref{fig:prfmnc:precieved:rank} shows participants' ranking for their desired patterns. A Friedman test confirmed that preferences differed significantly across the eight patterns ($\chi^2(7)=160.4$, $p<0.001$; Kendall's $W=0.23$). Holm-corrected Wilcoxon signed-rank post-hoc tests showed that \AlternativeFraming was preferred over every other pattern (all $p\leq0.009$), while \InterpretiveAlternative and \IntentionalMisleading were the least preferred.
The middle cluster (\SelectiveDistortion, \ScaffoldExplanation, \SocraticQuestioning, and \Oracle) was statistically indistinguishable. Notably, \Oracle was not preferred over the more contemplative patterns. Preference therefore does not track accuracy: \Oracle produced the largest accuracy gains yet ranked only mid-pack, while the most-preferred pattern, \AlternativeFraming, produced no accuracy gain that survived correction Therefore \textbf{H4} is not supported.


On average, \emph{Alternative framing} pattern was ranked as the most preferred. Participants explained that this rhetorical pattern \textit{``helped provide [them] with facts and context to the question in a quick and uncomplicated format, assisting [them] in figuring out the correct answers''} (P2)
Conversely, \InterpretiveAlternative and \IntentionalMisleading ranked worst on average, with most participants ranking them fifth or worse.
While participants recognized the misleading information of \IntentionalMisleading pattern and this pattern led them to \textit{``second guess what it said"} (P7), they did not prefer this pattern due to ``justifying the wrong answer'' (P6)

\InterpretiveAlternative was ranked the \textit{``worst one"} (P82) by participants due to \textit{``making [them] second guess and even more confused"} (P82)



Contrary to the observation that \TriggeringDistrust pattern helped participant accuracy, participants did not rank this pattern highly. P5 mentioned ``\textit{[it] provided direct answers, however, included information in the advice that was untrustworthy/incorrect. For example, saying that a baseball player played underwater ping pong for various MLB teams.}'' Generally, participants suggested patterns that had ``\emph{short and sweet}'' (P9), ``\emph{easy to understand}''(P13) and ``\emph{straightforward without having random thoughts unrelated to the claim}''(P60).

Participants mostly preferred to-the-point, concise patterns, short advice, and those who could help them complete the task as easily and quickly as possible.Participants especially disliked patterns who asked questions like \emph{Socratic Questioning} or \emph{Interpretive Alternate} patterns. For example, P49 said he preferred a rhetorical pattern with ``\emph{no additional questions, [which] answers the original question with no additional unrelated information, and no questions after.}'' 



\subsection{RQ2: How do rhetorical patterns influence participants' contemplation?}

In this section, we compare different patterns based on how they impact participants to contemplate and think critically.

\subsubsection{Time on Task}

\begin{figure*}
  \centering
    \subfloat[Total claim-evaluation time per pattern, derived from interaction timestamps. Time on task differed significantly across patterns ($\chi^2(7)=29.3$, $p<0.001$): \Oracle was the fastest, while \SocraticQuestioning and \InterpretiveAlternative took the longest.]{%
      \includegraphics[width=0.49\textwidth]{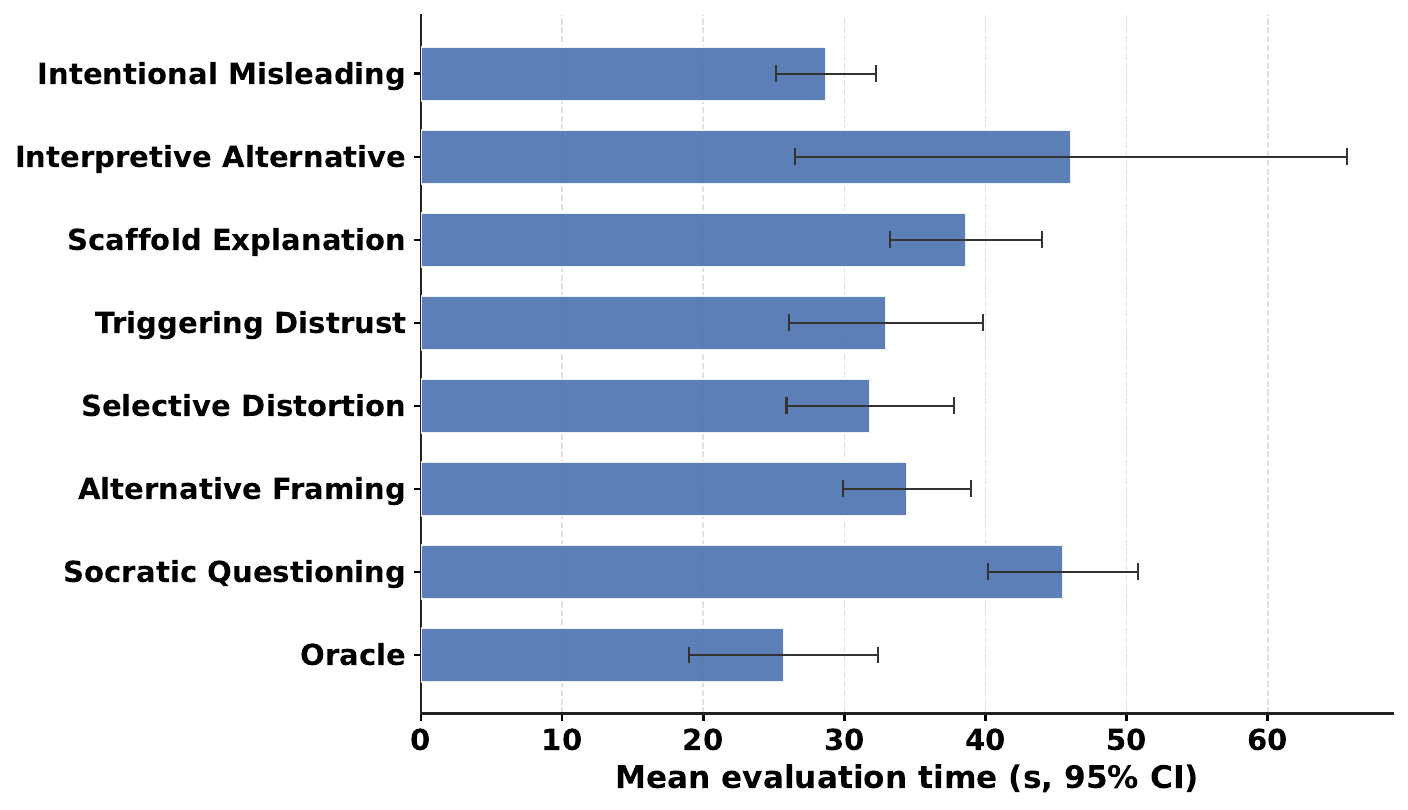}%
      \label{fig:frc:total_time}}
    \hfill
    \subfloat[Likert ratings (1--5) of how understandable and easy to use each pattern was; \Oracle scored highest on both and \InterpretiveAlternative lowest.]{%
      \includegraphics[width=0.49\textwidth]{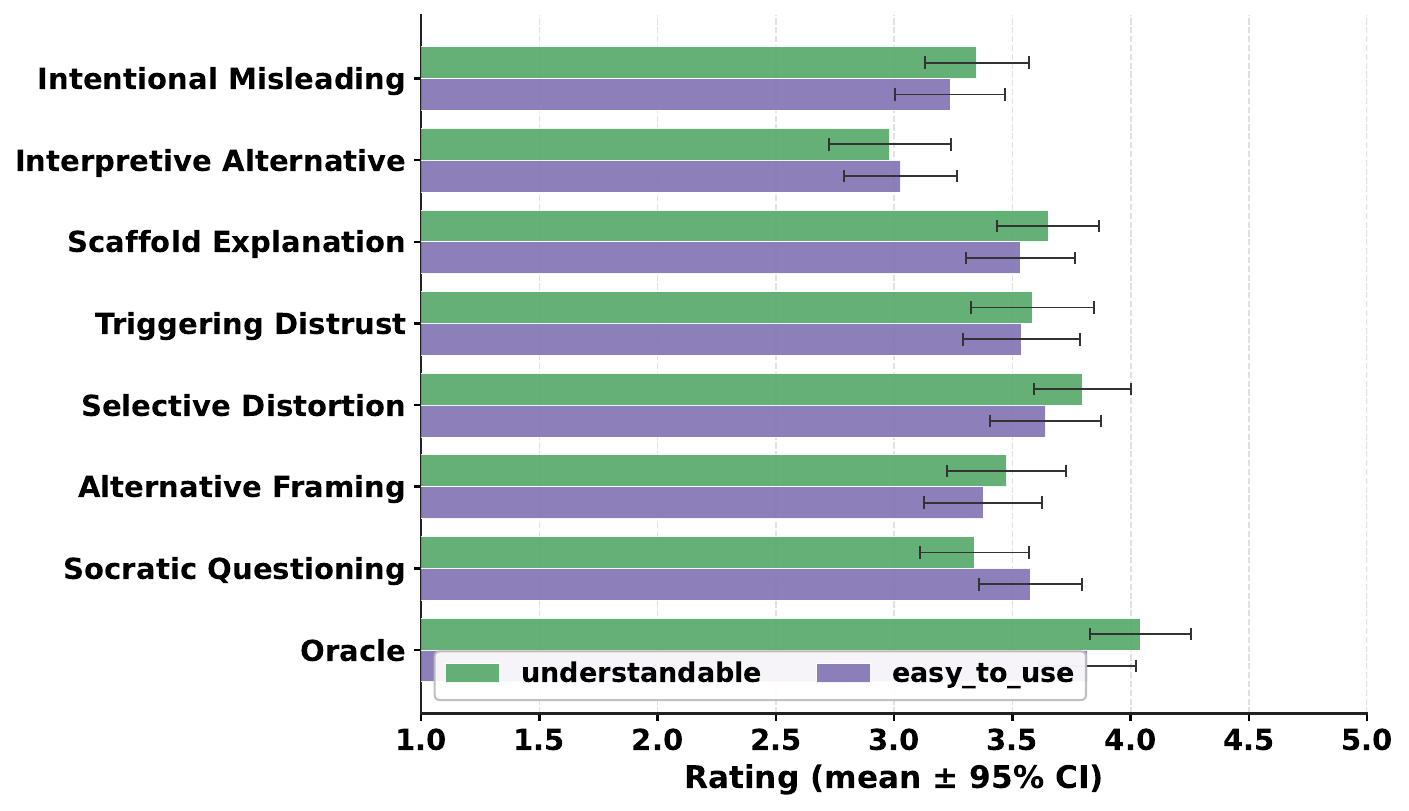}%
      \label{fig:frc:undrs_easy2use_easy2eval}}
  \caption{Task time (a) and usability ratings (b) across patterns.}
  \label{fig:frc}
\end{figure*}

In the literature on evaluating human problem-solving abilities, time on task or response time has been shown to be an indicator of critical thinking~\cite{Travers2016The,Alos-Ferrer2016Cognitive,Bilancini2024Manipulating}. We treat time on task as a behavioral proxy for effortful engagement rather than a direct measure of reflection, following prior work that uses task time as an indicator of effort in critical-thinking assessment~\cite{hyytinen2023self} and self-report as a complementary measure of reflective thinking~\cite{ghanizadeh2017interplay, kleemola2021exploring}. We measured each participant's total evaluation time for each claim and compared the average time across different advice patterns. Figure~\ref{fig:frc:total_time} shows the total claim evaluation time for each pattern.
The results show that \Oracle had the least time on task since it was \textit{``to the point''} (P8). However, claim evaluation with \SocraticQuestioning and \InterpretiveAlternative took the longest compared to other patterns. Participants reported that these patterns were \textit{more time consuming and didn't help [them] the way [they] would like''}(P41) and \textit{took [them] a while to understand it well''}(P7).

Time on task differed significantly across patterns ($\chi^2(7)=29.3$, $p<0.001$). Measured against \Oracle directly, however, only \SocraticQuestioning was significantly slower ($+18.3$s, $p<0.001$). \AlternativeFraming ($+7.3$s, $p=0.07$) and \InterpretiveAlternative ($+19.4$s, $p=0.11$) trended in the predicted direction but did not reach significance. \ScaffoldExplanation, which we did not classify as cognitively demanding, was also reliably slower than \Oracle ($+11.7$s, $p<0.01$). \textbf{H5} is therefore only partially supported: the pattern that most clearly imposes cognitive demand does slow users down, but our three-pattern grouping does not cleanly separate them from others.

Counterintuitively, this additional time did not translate into commensurate accuracy gains: neither \SocraticQuestioning nor \InterpretiveAlternative showed a significant improvement in claim accuracy after correction, despite demanding the most time, well below the gains of more concise patterns such as \Oracle and \ScaffoldExplanation. Time on task and accuracy therefore appear dissociated and hence \textbf{H6} is not supported. Similar dissociations between time on task and performance have been shown in tasks involving the Cognitive Reflection Test~\cite{Kramer2023Testing} and computer-based reading tasks~\cite{Goldhammer2014The}.


\subsubsection{Reflection}

Effective AI-assisted information systems should encourage users to engage in critical thinking rather than passive acceptance of recommendations. In our study, participants frequently reported that certain patterns prompted them to reflect more deeply on their evaluations. The \ScaffoldExplanation pattern appeared as particularly effective at initiating critical thinking by elaborating on the logical reasoning behind claims. As P45 noted, ``\emph{I think [it] made me think more because I had to read more information and come to a conclusion}.''

In addition, we observed that awareness of potential manipulation also triggered heightened scrutiny. When participants recognized that the \IntentionalMisleading pattern was \textit{``trying to confuse them'' }(P61), it actually \textit{``makes [them] think the claim could possibly be true''}(P61) through questioning.

When asked which patterns prompted the most reflection, participants mainly identified \Oracle, \ScaffoldExplanation, \SocraticQuestioning, and \InterpretiveAlternative. Participants attributed their increased reflection to two primary characteristics: (1) patterns who posed questions and (2) those who provided detailed responses. 

P46, elaborated on this noting:``\emph{\InterpretiveAlternative gave me a statement and then asked questions that made me doubt whether my answer was going to be correct. Very frustrating but I eventually stopped reading the doubtful questions and just relied on the first statement since those were correct info.}'' 

Similarly, P68 found the \ScaffoldExplanation pattern reflective because ``\emph{[it] gave [them] a ton of information and answered the question perfectly, making [them] change [their] answer}.'' Also, participants reported that  this pattern initiates this critical thinking in users by elaborating on the logical flow of the claim. For example, P45 said, ``\emph{I think [it] made me think more because I had to read more information and come to a conclusion}.''
We also found that if people understand that patterns are misleading them, they reflect more on the information's credibility. 
When \IntentionalMisleading tried to mislead P61 with a question, they said, ``\emph{[the] question at the end of the advice is like trying to confuse me or make me think it could possibly be true}.''



\subsubsection{Rhetorical Patterns Usability}


While inducing reflection is valuable, information systems should help users easily understand and apply the guidance provided. We asked participants to rate each rhetorical pattern using five-point Likert scales, from very little (1) to very high (5), across understandability (How easy was it to understand the information in the rhetorical pattern?) and usability (How easy to use was the information in the rhetorical pattern?).

As shown in Figure~\ref{fig:frc:undrs_easy2use_easy2eval}, participants rated the \Oracle pattern as the easiest to use. describing it as ``\emph{easy to use}'' [P6] and ``\emph{easy to utilize}'' [P15]. Conversely, the \InterpretiveAlternative pattern was rated most difficult to use and evaluate. Participants explained this rating by noting it was ``\emph{hard to determine if it was accurate or not}'' [P42] and that ``\emph{its large number of questions made it hardly useful}'' [P61].


As shown in Figure~\ref{fig:frc:undrs_easy2use_easy2eval}, participants rated \Oracle and \SelectiveDistortion as both the most understandable and the easiest to use and evaluate. P68 illustrated this connection when discussing the \SelectiveDistortion pattern: ``\emph{But the advice it gave was simple to understand and when it was correct it was very quick and easy}.'' Similarly, participants praised the \emph{Oracle} pattern for being ``\emph{easy to use}'' [P6] and ``\emph{easy to utilize}'' [P15], reflecting how comprehensibility translates directly into practical utility in claim evaluation tasks.

Conversely, \InterpretiveAlternative was rated lowest across all usability dimensions—comprehensibility, ease of use, and ease of evaluation. This pattern's complexity manifested as both difficulty in understanding and difficulty in application. P93 described how lack of clarity undermined usability: ``\emph{The pattern would report advice but then ask many questions after making it confusing to understand/trust the advice}.'' P42 and P61 reinforced this pattern, noting it was ``\emph{hard to determine if it was accurate or not}'' and that ``\emph{its large number of questions made it hardly useful},'' respectively.

These findings surface a critical design challenge for Information agents: How can we design patterns that encourage users to engage in critical thinking while maintaining the clarity necessary for users to understand and apply the guidance provided?



\subsubsection{The Costs of Excessive Friction}

While friction can promote contemplation, excessive friction imposes mental costs on users. Participants' frustration correlated strongly with both time consumption and confusion. P80's response to the \InterpretiveAlternative pattern illustrates this tension: ``\emph{These advises were very confusing. The questions at the end of the advice could go either way as far as true or false, and that was really hard to use to determine the validity of the statement. I did not like this form of advice, although it seems I got more statements correct with this pattern}.'' This frustration was particularly pronounced for patterns that employed questioning or challenged users' initial judgments. P25 expressed similar concerns about \IntentionalMisleading: ``\emph{The pattern asking a question at the end of its response added confusion to how much I trusted its initial answers}.'' Furthermore, some participants percived patterns that induced less reflection as more trustworthy and enjoyable to use. P90 praised the \ScaffoldExplanation pattern, stating ``\emph{[it] was very straightforward, easier to understand and trustworthy and the experience was great and interesting, and it was also very quick in answers, and I will likely use this in the future}.'' 


Our findings reveal that designing effective patterns with appropriate friction represents an optimization challenge: balancing the benefits of induced reflection against the costs of user frustration. patterns operating at moderate levels of friction and frustration, such as \TriggeringDistrust, appear to achieve a more favorable trade-off than those at either extreme.

\section{Discussion}
\label{sec:discussion}

\subsection{Implications for AI Design}
Our design explorations revealed that the default rhetoric pattern of AI assistants, i.e., delivering direct responses, may inadvertently suppress users' critical thinking, a finding that aligns with prior work showing that LLMs affect user confidence independently of accuracy~\cite{zhang2020effect, bansal2021does} and that users frequently over-rely on automated advice even when it is incorrect~\cite{cummings2017automation}. Advisory styles that introduce deliberate friction offer one design path toward counteracting this tendency. However, friction must be carefully designed. Styles like \InterpretiveAlternative, which illustrate alternatives and raise questions without providing grounding, increased time on task and frustration while yielding no significant accuracy gain, far below those of more concise, grounded patterns. This points to a core design principle: friction should be \textit{legible}, meaning users should understand why they are being asked to pause and reconsider, and the path forward should remain clear. This resonates with HCI research showing that small interruptions can shift users toward more reflective engagement without substantially degrading task performance~\cite{cox2016design, chen2024positive}, and that selective frictions in LLM contexts can reduce over-reliance while keeping accuracy roughly stable~\cite{deschryver2025friction, inan2025positive}. Critically, poorly targeted friction produces frustration rather than reflection~\cite{deschryver2025friction, cox2016design}, reinforcing that legibility is a condition for friction to function as intended.

Perhaps the most striking finding is that adversarial conditions, patterns designed to mislead or distort, were associated with accuracy no worse than \Oracle. One reading is inoculation theory and devil's advocacy~\cite{schwenk1990effects}, in which detecting a misleading advisor which triggers motivated scrutiny that extends to the claim itself. We advance this reading only tentatively: because we ran no manipulation check, we cannot verify that participants perceived the distortion as distortion, and a more parsimonious account is response bias.

These findings together suggest that no single rhetorical style is universally optimal, pointing toward systems that dynamically adapt their pattern based on user behavior and task context. Objective factual claims may call for direct, low-friction styles, while subjective or high-stakes topics may benefit from more contemplative patterns that resist premature closure. LLM post-training interventions could steer systems toward 
this adaptive behavior, prioritizing reflection-inducing styles 
in higher-risk contexts.

\subsection{The Accuracy-Preference Conflict}
One of the most practically consequential findings in our study is that what users prefer and what actually helps them are often in conflict. \AlternativeFraming was the most preferred advisor, with participants describing it as ``\emph{concise and easy to use,}'' yet its accuracy gain was not significant. \ScaffoldExplanation, which generated the highest accuracy improvements and the deepest reflection, ranked third in preference. Notably, the directive \Oracle was not preferred over the more contemplative patterns despite its accuracy advantage. This divergence poses a fundamental design question for deployed systems. If an AI assistant optimizes for user satisfaction, it will likely converge toward brief, direct, confident responses, precisely the style most associated with over-reliance and miscalibration~\cite{zhang2020effect, cummings2017automation}. This risk is real: sycophancy and anthropomorphism have been documented as emergent failure modes in LLMs trained to maximize user preference, where models learn to tell users what they want to hear rather than what is accurate~\cite{ford2023towards}, and to adopt social personas that encourage trust beyond what is warranted~\cite{cheng2025social}.

This tension also explains why \Oracle's accuracy advantage does not settle the design question. An oracle presupposes an authoritative, gold-standard source, which is rarely available in the real world, precisely for the novel, breaking, or contested claims where verification matters most. The contemplative patterns therefore remain the relevant design space despite \Oracle's edge in accuracy, since they must carry the load exactly where an oracle cannot be constructed.

Our findings suggest that rhetorical design faces the same trap: optimizing for what users prefer in the moment may systematically undermine the accuracy and reflection that information evaluation demands. If it prioritizes accuracy or reflection, it risks alienating users who find the interaction effortful or frustrating, thereby reducing engagement over time. We suggest that effective advisory systems should treat this as an explicit optimization target, potentially surfacing the trade-off transparently to users and allowing them to choose their preferred balance between ease and depth depending on the stakes of the decision at hand.

\subsection{Ethical Considerations}
Deploying adversarial rhetorical patterns in real systems raises ethical questions that our study does not fully resolve. Several conditions in our study, namely \IntentionalMisleading, \TriggeringDistrust, and \SelectiveDistortion, were by design deceptive, and participants were not informed of this during the task. While our IRB-approved debriefing addressed this in the study context, the implications for deployed systems are more fraught. Even if adversarial patterns improve accuracy on average, they do so by manipulating users without their knowledge or consent. Users who eventually recognize the manipulation, as several participants did, may lose trust not only in the specific advisor but in AI-assisted information tools more broadly. We therefore caution against deploying adversarial patterns in production systems without explicit user awareness and consent. A more ethically grounded approach might be to disclose the advisory style to users, for instance by framing \SocraticQuestioning as a ``reflection mode'' that the user can opt into, preserving the accuracy benefits of deliberate friction while respecting user autonomy. Future work should examine whether disclosing the rhetorical 
strategy to users preserves or diminishes its effectiveness.

\section{Limitations}
\label{sec:limitations}

Our study has several limitations. Each 
pattern was evaluated on only five manually crafted claims in 
a single-turn setting, which restricts generalizability across 
claim types and interaction contexts. Real-world advisory 
interactions are rarely one-shot, and rhetorical 
effects may differ across multi-turn dialogues 
or more subjective, high-stakes topics. Future work should 
validate these findings with longer interaction histories and 
LLM-generated advice each pattern, which would 
also reduce the author-specific confounds of manual 
construction. Our within-subject design introduces potential 
fatigue and skepticism effects across the 40-claim session, 
though randomized advisor assignment mitigates 
ordering bias. Our Prolific sample also skews toward educated, 
digitally literate users; future studies should 
recruit populations more vulnerable to misinformation, such as 
older adults or those with lower media literacy, to test whether 
our accuracy and preference findings hold more broadly.
Several adversarial conditions varied both rhetorical framing and information completeness, making it difficult to isolate rhetoric as the sole active variable; future work should address this by holding information completeness constant.
This confound is most acute for the adversarial accuracy gains, where detected manipulation and missing information may each independently drive the observed effect.

Because \SelectiveDistortion and \TriggeringDistrust alter the information itself rather than only its wording, we accordingly limited our claims about rhetoric to the patterns that primarily vary phrasing.
Our claim set is also unbalanced by truth value (14 of 40 claims are true), and truth value is fixed per pattern rather than rotated across participants. Since ``false'' is the correct verdict for 65\% of claims, a pattern that nudges participants toward skepticism can raise measured accuracy without improving their ability to discriminate true from false. This response bias is a plausible alternative account of why the distorting patterns performed no worse than \Oracle, and future work should balance the claim set and report accuracy separately by claim truth value.

Finally, the absence of manipulation checks means we cannot verify whether participants processed each pattern as intended, so we framed our interpretations as associations rather than causal claims,
a gap future work should address with brief post-condition 
checks.

\section{Conclusion}
We investigated how rhetorical patterns in AI advisory responses shape user reasoning, accuracy, and reflection in a fact verification task. Across eight patterns, how AI systems communicate mattered as much as what they communicate: \ScaffoldExplanation and \Oracle produced the strongest accuracy gains, adversarial patterns improved accuracy through motivated scrutiny, and user preference consistently diverged from user performance. These findings challenge the assumption that direct, confident responses are always best, and suggest that deliberate friction can be a productive resource in AI-assisted information evaluation. We hope this work encourages designers and researchers to treat rhetorical style as a first-class design variable in conversational AI systems.

\section*{Acknowledgments}
This research is supported by the U.S. Defense Advanced Research
Projects Agency (DARPA) under transaction
award HR00112490374 from the Friction for
Accountability in Conversational Transactions
(FACT) program. Any opinions, findings, conclusions, or recommendations expressed here are
those of the authors and do not necessarily reflect
the view of the sponsors.

\bibliography{main}
\bibliographystyle{IEEEtran}

\end{document}